
\input harvmac.tex
\input epsf.tex
\parindent=0pt
\parskip=5pt

\def\IR{{\hbox{{\rm I}\kern-.2em\hbox{\rm R}}}}
\def\IB{{\hbox{{\rm I}\kern-.2em\hbox{\rm B}}}}
\def\IN{{\hbox{{\rm I}\kern-.2em\hbox{\rm N}}}}
\def\IC{{\hbox{{\rm I}\kern-.6em\hbox{\bf C}}}}
\def\IZ{{\hbox{{\rm Z}\kern-.4em\hbox{\rm Z}}}}
\def\IP{{\hbox{{\rm I}\kern-.2em\hbox{\rm P}}}}

\def\Z{{\bf Z}}
\def\IZ{\Z}

\noblackbox

\Title{\vbox{\baselineskip12pt
\hbox{NSF-ITP-96-55}
\hbox{hep-th/9606176}}}
{Multiple Realisations of ${\cal N}{=}1$  Vacua in Six Dimensions}

\centerline{ \bf Eric G. Gimon$^a$ and Clifford V. Johnson$^b$}
\bigskip\bigskip\centerline{{\it 
$^a$Department of Physics / $^b$Institute for Theoretical Physics,}}
\centerline{{\it University of California,}}
\centerline{{\it Santa Barbara, CA~93106, USA }}
\footnote{}{\tt $^a$egimon@physics.ucsb.edu, $^b$cvj@itp.ucsb.edu}
\vskip1.4cm
\centerline{\bf Abstract}
\vskip0.7cm
\vbox{\narrower\baselineskip=12pt\noindent
A while ago, examples of ${\cal N}{=}1$ vacua in $D{=}6$ were
constructed as orientifolds of Type~IIB string theory compactified on
the $K3$ surface. Among the interesting features of those models was
the presence of D5--branes behaving like small instantons, and the
appearance of extra tensor multiplets. These are both
non--perturbative phenomena from the point of view of Heterotic string
theory. Although the orientifold models are a natural setting in which
to study these non--perturbative Heterotic string phenomena, it is
interesting and instructive to explore how such vacua are realised in
Heterotic string theory, M--theory and F--theory, and consider the
relations between them. In particular, we consider models of M--theory
compactified on $K3{\times}S^1/\Z_2$ with fivebranes present on the
interval. There is a family of such models which yields the same
spectra as a subfamily of the orientifold models. By further
compactifying on $T^2$ to four dimensions we relate them to
Heterotic string spectra. We then use Heterotic/Type~IIA duality to
deduce the existence of Calabi--Yau 3--folds which should yield the
original six dimensional orientifold spectra if we use them to
compactify F--theory. Finally, we show in detail how to take a limit
of such an F--theory compactification which returns us to the Type~IIB
orientifold models. }

\vskip0.5cm

\Date{June, 1996}

\baselineskip13pt

\lref\ericjoe{E. G. Gimon and J. Polchinski, {\sl `Consistency 
Conditions for Orientifolds and D--Manifolds'}, hep-th/9601038.}

\lref\atishi{A. Dabholkar and  J. Park, {\sl `An Orientifold of 
Type--IIB Theory on $K3$'}, hep-th/9602030.}

\lref\atishii{A. Dabholkar and  J. Park, {\sl `Strings on Orientifolds'}, 
hep-th/9604178.}

\lref\joe{J. Polchinski, {\sl `Tensors from $K3$ Orientifolds'}, 
hep-th/9606165.}

\lref\gojoe{J. Polchinski, {\sl `Dirichlet Branes and Ramond--Ramond
Charges in String Theory.'}, Phys. Rev. Lett. {\bf 75} (1995) 4724,
hep-th/9510017.}

\lref\dnotes{J. Polchinski, S. Chaudhuri and C. V. Johnson, {\sl 
`Notes on D--Branes'}, hep-th/9602052.}

\lref\stringstring{M. J. Duff and R. R. Khuri, 
Nucl. Phys. {\bf B411} (1994) 473\semi
M. J. Duff and R. Minasian, Nucl. Phys. {\bf B436} (1995) 507.}

\lref\ed{E.~Witten, {\sl `String Theory Dynamics in Various
 Dimensions'}, 
Nucl. Phys. {\bf B443} (1995) 85, hep-th/9503124.}

\lref\edjoe{J. Polchinski and E. Witten {\sl `Evidence for Heterotic/Type~I
 Duality'},
preprint Nucl. Phys. {\bf B460}  (1996) 525, hep-th/9510169.}
\lref\sagnotti{A. Sagnotti, {\sl `A Note on the
 Green--Schwarz Mechanism in Open String Theories'},
 Phys. Lett. {\bf B294} (1992) 196, hep-th/9210127.} 
\lref\solitons{A. Strominger {\sl `Heterotic Solitons'},
 Nucl. Phys. {\bf B343} (1990) 167\semi
C. G. Callan, J. A. Harvey and A. Strominger, {\sl `Solitons 
in String
Theory',}, Trieste Notes, World Scientific (1991), hep-th/9112030.}

\lref\small{E. Witten, {\sl `Small Instantons in String
Theory',} preprint IASSNS-HEP-95-87, hep-th/9511030.}
\lref\joetc{M. Berkooz, R. G. Leigh, J. Polchinski, J. H. Schwarz, N. 
Seiberg and E. Witten, {\sl `Anomalies, Dualities,  and 
Topology of D=6 N=1 Superstring Vacua'}, hep-th/9605184.}

\lref\eguchi{T. Eguchi and A. J. Hanson, {\sl `Asymptotically 
Flat Self--Dual Solutions to Euclidean Gravity'}, Phys. Lett. {\bf B74} 
(1978), 249.}

\lref\green{M. B. Green and J. H. Schwarz, {\sl `Anomaly 
Cancellation In Supersymmetric $D{=}10$ Gauge Theory and Superstring Theory'},
  Phys. Lett. {\bf B149} (1984) 117.}
\lref\edcomm{E. Witten, {\sl `Some Comments on String Dynamics'}, 
hep-th/9507121.}
\lref\hethet{M. J. Duff, R. Minasian and E. Witten,
{\sl `Evidence for Heterotic/Heterotic Duality'}, hep-th/9601036.}
\lref\duffhethet{M. J. Duff and J. X. Lu, {\sl `Loop Expansions and 
String/Fivebrane Duality'}, Nucl. Phys. {\bf B357} (1991) 534\semi
M. J. Duff and R. R. Khuri, {\sl `Four--Dimensional String/String
Duality'}, Nucl. Phys. {\bf B411} (1994) 473,
 hep-th/9305142\semi M. J. Duff and
R. Minasian, {\sl `Putting String/String Duality to the Test'},
Nucl. Phys. {\bf B436} (1994) 507, hep-th/9406198}
\lref\edfive{E. Witten, {\sl `Fivebranes and M--Theory on an Orbifold'}, 
hep-th/9512219.}
\lref\vafaF{C. Vafa, {\sl `Evidence for F--Theory'}, hep-th/9602022.}
\lref\vafamorrison{D. R. Morrison and C. Vafa, {\sl `Compactifications of
 F--Theory on Calabi--Yau Manifolds I $\&$ II'}, hep-th/9602114 $\&$ 
hep-th/9603161.}
\lref\font{G. Aldazabal, A. Font, L. E. Ib\'a\~nez and F. Quevedo, {\sl 
`Heterotic/Heterotic Duality in $D{=}6,4$'}, hep-th/9602097.}
\lref\klemm{A. Klemm, W. Lerche and P. Mayr, {\sl `$K3$--Fibrations and 
Heterotic--Type~II String Duality'}, hep-th/9506112.}
\lref\ericme{E. G. Gimon and C. V. Johnson, {\sl `$K3$ Orientifolds'}, 
hep-th/9604129.}
\lref\petred{P. Ho\u rava and E. Witten, {\sl `Heterotic and Type~I 
String Dynamics from Eleven Dimensions'},  Nucl. Phys. {\bf B460}  
 (1996) 506, hep-th/9510209
}
\lref\shamit{S. Kachru and C. Vafa, {\sl `Exact Results for $N{=}2$ 
Compactifications of Heterotic Strings'}, Nucl. Phys. {\bf B450} (1995) 
69, hep-th/9505105. }
\lref\senF{A. Sen, {\sl `F--Theory and Orientifolds '}, hep-th/9605150.}
\lref\senM{A. Sen, {\sl `M-Theory on $(K3{\times}S^1)/\Z_2$'}, 
hep-th/9602010.}
\lref\edphase{E. Witten, {\sl `Phase Transitions in M--Theory and 
F--Theory'},
 hep-th/9603150\semi N. Seiberg and E. Witten, {\sl `Comments on String 
Dynamics in Six Dimensions'}, hep-th/9603003.}
\lref\duffphase{M. J. Duff, H. Lu and C. N. Pope, {\sl `Heterotic Phase 
Transitions and Singularities of the Gauge Dyonic String'}, hep-th/9603037.}
\lref\ganor{O. J. Ganor, A. Hanany, {\sl `Small $E_8$ Instantons and 
Tensionless Non--critical Strings'}, hep-th/9602120.}
\lref\schwarz{M. B. Green, J. H. Schwarz and P. C.  West, 
{\sl `Anomaly Free Chiral Theories in Six Dimensions'},
Nucl. Phys. {\bf B254} (1985) 327. }
\lref\sunil{K. Dasgupta and S. Mukhi, {\sl `Orbifolds of M--theory'},
 hep-th/9512196.}
\lref\philip{P. Candelas, E. Perevalov and  G. Rajesh, {\sl `F--Theory
 Duals of Nonperturbative Heterotic $E_8{\times}E_8$ Vacua in 
Six Dimensions'}, hep-th/9606133.}
\lref\sodual{E.~Witten, {\sl `String Theory Dynamics in Various
 Dimensions'}, 
Nucl. Phys. {\bf B443} (1995) 85, hep-th/9503124.}

\newsec{Introduction}
In six dimensions, vacua with ${\cal N}{=}1$ supersymmetry have a rich
and interesting structure. Due to potential chiral anomalies, such
vacua are subject to constraints which enable us learn a great deal
about the nature of the various sibling string theories (and their
parent theories) which can give rise to them.

Due to various perturbative and non--perturbative symmetries, a given
spectrum may be obtained in many different ways.  For example, by
studying $SO(32)$ heterotic string vacua in $D{=}6$, using constraints
from the chiral anomalies, the presence of non--perturbative effects
attributable to small instantons may be deduced\small. In the dual
type~I theory, these non--perturbative effects can be studied
perturbatively as D5--branes\refs{\small,\ericjoe}. This is a
consequence of ten dimensional strong/weak coupling duality between
the two $SO(32)$ theories\sodual.

Another example is heterotic/heterotic duality\hethet. The conjecture
about the existence of this duality was motivated\duffhethet\ in part
by the structure of the factorised 8--form polynomial associated to
the anomaly of a $D{=}6$ heterotic $K3$
compactification\foot{Amusingly, it seems that it was almost rejected
for the same reason, as the signs of some of the $\Tr F^2$ terms in
one of the factors (for the standard embedding) signaled that there
would be unpleasant behaviour somewhere in coupling space. This is now
interpreted as the sign of a phase transition\hethet.}. The
realisation of the possibility of non--perturbative gauge groups due
to small instantons allowed the conjecture to be confirmed\hethet\ in
terms of a $K3$ compactification of the $E_8{\times} E_8$ heterotic
string with a choice of vacuum gauge bundle with 12 instantons
assigned to each $E_8$. The conjectured duality map acts
non--trivially on the gauge group and hypermultiplets to exchange the
perturbative and non--perturbative contributions to this $D{=}6$
string vacuum.

In an apparently (at the time) different setting, an orientifold
construction of type~I string theory compactified on $K3$ to six
dimensions was presented\ericjoe.  The $K3$ manifold was in its
$T^4/{\bf Z}_2$ orbifold limit. Upon closer examination, a relation
between (a special case of) that model and the heterotic/heterotic
construction may be deduced\joetc: This model, constructed using
perturbative string techniques, enjoys the presence of two
isomorphic sectors to its gauge group (and charged hypermultiplets),
attributable to the necessary presence of both D9-- and D5--branes.
The isomorphism between the two sectors is simply realised as
perturbative T--duality in the $X^6,X^7,X^8$ and $X^9$ directions of
the $K3$ torus (denoted $T_{6789}$--duality henceforth).

As this model is a compactification of $SO(32)$ type~I string theory,
it ought to be strong/weak coupling dual to a compactified heterotic
model, by virtue of their relationship\sodual\ in $D{=}10$. This turns
out to be true, and the details are very instructive. It is the
$E_8{\times}E_8$ $(12,12)$ compactification which turns out to be
relevant, with the subtlety residing in the fact that what were
naively small $SO(32)$ instantons from the type~I perspective, or
small $E_8$ instantons from the heterotic perspective, are actually
properly thought of as ${\sl spin}(32)/\IZ_2$ instantons for that
particular embedding\joetc. Under this particular type~I/heterotic
map, the perturbative $T_{6789}$--duality of the type~I string induces
the conjectured heterotic/heterotic strong/weak coupling duality map
in the heterotic picture. Along the way, we learn again that the
distinction between the two types of heterotic string is weakened when
we leave ten dimensions, this time compactifying on $K3$.

This heterotic model also has a realisation in F--theory. This twelve
dimensional setting, which may be regarded as a powerful means of
generating new consistent backgrounds for the type~IIB string, gives
rise to ${\cal N}{=}1$ vacua in $D{=}6$ after compactification on an
elliptically fibred Calabi-Yau three--fold\vafaF. In fact, it has been
shown that the family of Calabi--Yau manifolds which may be described
as a fibration of a torus over the `ruled surfaces' $F_n$, are the
appropriate elliptic 3--folds on which to compactify F--theory in
order to realize duals to the heterotic vacua obtained by
compactifying on $K3$ with instanton embedding $(12{-}n,12{+}n)$. In
the case $n{=}0$, the surface $F_0$ is simply the product of
two--spheres $\IP_1{\times}\IP_1$. The elliptic 3--fold is the
ubiquitous degree 24 hypersurface in weighted projective space ${\rm
W}\IP(1,1,2,8,12)$, denoted $X_{24}(1,1,2,8,12)$. The
heterotic/heterotic duality map of the (12,12) model\foot{The (10,14)
model turns out to be the same model, as conjectured in ref.\font\ and
demonstrated in the F-theory context in ref.\vafamorrison. This is
simply because the relevant elliptic fibration over $F_2$ is
isomorphic to the one over $F_0$.}\ becomes the exchange of the two
$\IP_1$'s, one carrying the data to be labeled as perturbative in the
heterotic model and the other $\IP_1$ carrying the non--perturbative
data. As for the orientifold setting, this is another arena in which
the perturbative and non--perturbative structures (from the point of
view of the heterotic string) are treated on the same footing.  We
shall study this particular model some more in section~6, establishing
a direct relation between the orientifold and F--theory constructions,
following the ideas in  ref.\senF.

The purpose of this paper is to try to understand more of such
specific examples of ${\cal N}{=}1$ vacua in six dimensions in many
different settings. To this end, we shall follow a circular chain of
dualities studying special cases of some of the orientifold models
presented in refs.\refs{\ericme,\atishii}\ which are closely related to
the orientifold model discussed above.

We start by considering M--Theory.
The strong coupling limit of ten dimensional $E_8{\times}E_8$
heterotic string theory is M--theory on the orbifold $S^1/\Z_2$.  It
is (arguably) unnecessary to go to this eleven dimensional theory to
explain even the non--perturbative aspects of the $(12,12)=(10,14)$
$E_8{\times}E_8$ heterotic models, as they only require an appeal to
small ${\sl spin}(32)/\IZ_2$ instantons and not small $E_8$
instantons.  This will not remain true for the models we consider
here.  Specifically it is in M--theory that the non--perturbative
appearance of extra tensor multiplets in the six dimensional spectra
have a most natural description\edfive, and this shall be the starting
point for this paper's study (section~3) when we come to try to learn
more about the orientifold models we consider (recalled in section~2).

From the M--theory and heterotic string discussions in six dimensions,
we compactify further to four (in section~4), where in that ${\cal
N}{=}2$ context, the two theories are indistinguishable from each
other, as the extra tensors give rise to vectors in four dimensions.
Next, we use heterotic/type~IIA strong/weak coupling duality to take
us to type~II string theory, by conjecturing the existence of two new
(to us, at least) $K3$ fibred Calabi--Yau 3--folds suggested by the
spectra.

This leads us to consider returning to type~IIB theory (in section~5),
and we do so by going to F--theory, asking that we can compactify it
on the 3--folds. If they have an elliptic
fibration as well, we can construct six dimensional
${\cal N}{=}1$ vacua  yielding  the spectra we first started with.

To close the circle, we should be able to find a limit of F--theory
which directly gives a perturbative type~IIB background, and indeed
there is one.  Extending the ideas in ref.\senF\ in section~6 we show
how to recover from F--theory orientifold models which are simple
$T$--duals of the ones we started with.

Our circular route  is summarised below:

\vskip2.0cm
\hskip2.5cm\epsfxsize=3.0in\epsfbox{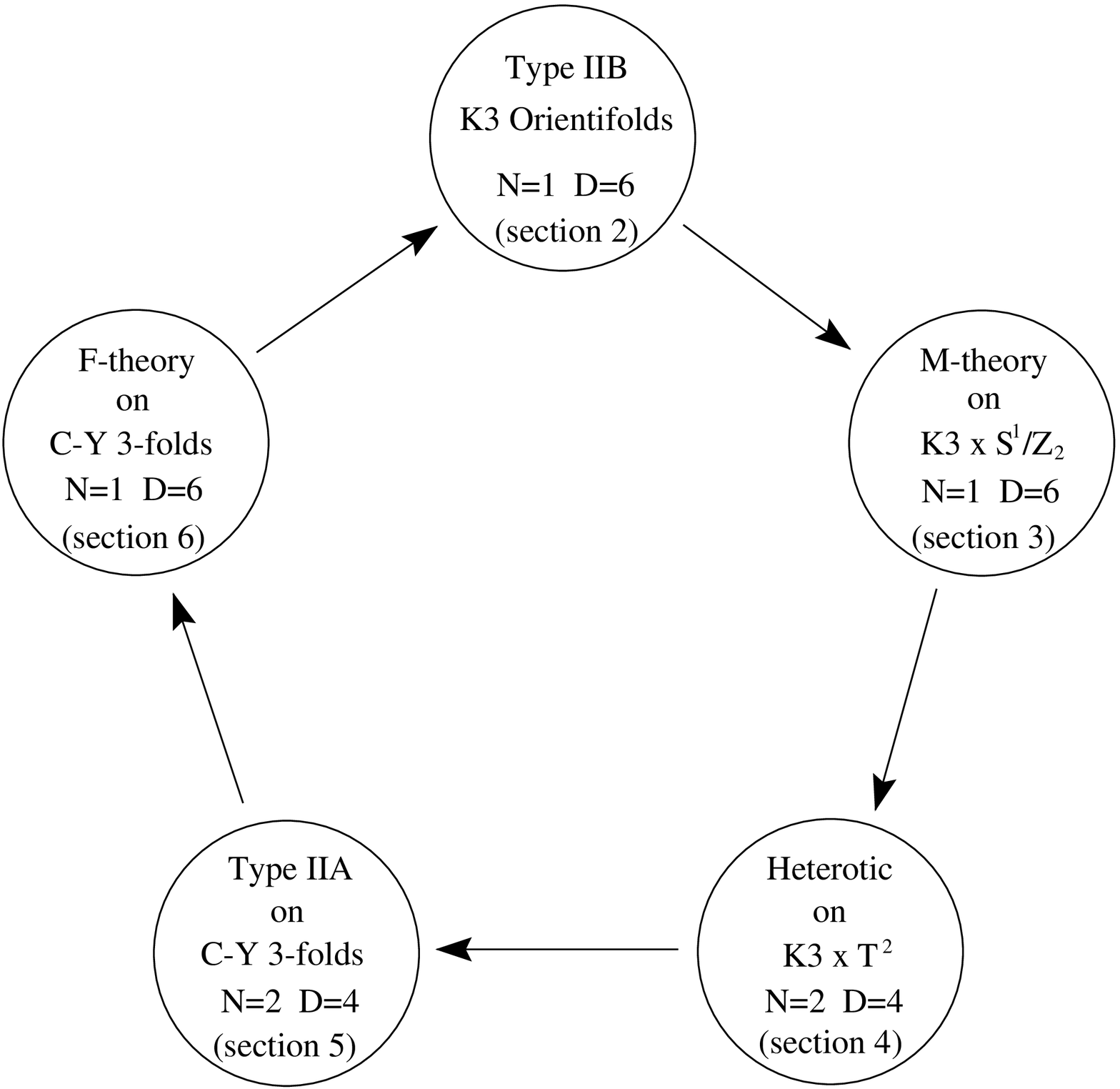}
\vskip1.0cm

There are three specific models which we take around the path we just
described. They are all quite different models in the orientifold
context in which they were first presented\foot{The first example
(originating in ref.\ericjoe) has been taken through most of this
discussion previously, as mentioned above. However, the details of the
direct F--theory connection are worked out here for the first
time.}\ericme\ericjoe. Once we get to M--theory we see that they are
probably related by rather simple phase transitions of the nature
discussed in refs.\refs{\ganor,\edphase,\duffphase}.  From our
comments about F--theory, they are likely to be closely related in
that context, although the issues are clouded somewhat by our
ignorance about the Calabi--Yau 3--folds which we conjecture to
exist. As a result, we complete the discussion of the connection of
F--theory with the orientifold models in detail only for the one
related to the heterotic/heterotic model, leaving the details of the
other models for the time when we have more data on the Calabi--Yau
3--folds.

\vfill\eject

\newsec{The Orientifold Models}
Let us briefly recall the models presented in ref.\ericme. They are
constructed as two types of orientifold of type~IIB strings which give
rise to six dimensional ${\cal N}{=}1$ supersymmetric models, with the
compact internal spacetime being the torus $T^4$ with orbifold
identifications.  In the terminology of ref.\ericme, type `$A$' models
contain the explicit appearance of the world--sheet parity operator
$\Omega$, and results in tadpoles which must be cancelled with
D9--branes. Type `$B$' models do not contain $\Omega$.  The worldsheet
operation $\Omega$ is consistently (at the level of perturbation
theory\joetc) combined with discrete spacetime rotation symmetries,
$\IZ_N$ which on their own, in a closed string setting, would have
resulted in an orbifold. The combination of the two results in an
`orientifold', and requires the addition of open string sectors,
personified in the form of `D--branes' to cancel the resulting
tadpoles. Depending upon the order of the rotation symmetry used for
the spacetime symmetry, there may be the requirement to introduce
D5--branes for this tadpole cancellation, with their world volumes
aligned with the non--compact directions $\{X^0,\ldots,X^5\}$.

For the particular choice of realisation of the symmetries $\IZ_N$ and
$\Omega$ used in the work of ref.\ericme, the models which arose
(denoted\foot{The $\IZ_2^A$ model was constructed in ref.\ericjoe, and
a special case of it is the dual of the (12,12) heterotic model, as
discussed above and later.}\ $\IZ_3^A,\IZ_4^A,\IZ_4^B,\IZ_6^A,\IZ_6^B$ and
$\IZ_2^A$) had the following closed string ${\cal N}{=}1$ $D{=}6$
spectra\foot{Throughout the paper, due to the presence of
supersymmetry it is enough to display only the bosonic part of the
spectra we consider.}\ (in addition to the usual supergravity
multiplet and tensor multiplet):

\bigskip
\vbox{
$$\vbox{\offinterlineskip
\hrule height 1.1pt
\halign{&\vrule width 1.1pt#
&\strut\quad#\hfil\quad&
\vrule#
&\strut\quad#\hfil\quad&
\vrule#
&\strut\quad#\hfil\quad&
\vrule width 1.1pt#\cr
height3pt
&\omit&
&\omit&
&\omit&
\cr
&\hfil Model&
&\hfil \vbox{\hbox{\hskip0.7cm Neutral}\vskip3pt\hbox{Hypermultiplets}}&
&\hfil \vbox{\hbox{\hskip0.9cm Extra}\vskip3pt\hbox{Tensor Multiplets}}&
\cr
height3pt
&\omit&
&\omit&
&\omit&
\cr
\noalign{\hrule height 1.1pt}
height3pt
&\omit&
&\omit&
&\omit&
\cr
&\hfil $\IZ_2^A$&
&\hfil 20&
&\hfil 0&
\cr
height3pt
&\omit&
&\omit&
&\omit&
\cr
\noalign{\hrule}
height3pt
&\omit&
&\omit&
&\omit&
\cr
&\hfil $\IZ_3^A$&
&\hfil 11&
&\hfil 9&
\cr 
height3pt 
&\omit& 
&\omit& 
&\omit&
\cr
\noalign{\hrule }
height3pt 
&\omit& 
&\omit& 
&\omit&
\cr
&\hfil $\IZ_4^A$&
&\hfil 16&
&\hfil 4&
\cr
height3pt 
&\omit& 
&\omit& 
&\omit&
\cr
\noalign{\hrule } 
height3pt 
&\omit& 
&\omit& 
&\omit&
\cr
&\hfil $\IZ_6^A$&
&\hfil 14&
&\hfil 6&
\cr 
height3pt 
&\omit& 
&\omit& 
&\omit&
\cr
\noalign{\hrule } 
height3pt
&\omit& 
&\omit& 
&\omit&
\cr
&\hfil $\IZ_4^B$&
&\hfil 12&
&\hfil 8&
\cr
height3pt 
&\omit& 
&\omit& 
&\omit&
\cr
\noalign{\hrule } 
height3pt 
&\omit& 
&\omit& 
&\omit&
\cr
&\hfil $\IZ_6^B$&
&\hfil 11&
&\hfil 9&\cr
}
\hrule height 1.1pt}
$$
}

With the exception of the $\IZ_2^A$ model, which is the model of
ref.\ericjoe, all of the models have extra self--dual tensor
multiplets (the perturbative heterotic stirng has only one), and a
reduced (from the standard 20) number of hypermultiplets corresponding
to the gravitational moduli of the $K3$ surface upon which we have
compactified.

In view of this, we ought not to think of these models with extra
tensors as compactifications of the type~I string on some $K3$
manifold, as we will run into trouble for many related reasons, some
of which we list below:

\item\item{{\bf 1.}} The origin of the extra tensors would be 
problematic in that setting. This is simply because geometrically they
would need to have a ten dimensional origin in a self--dual 4--form
$A_4$ of the R-R sector of the theory, as this is the only object
which could be contracted with the two--cycles of $K3$ to give
self--dual 2--forms in six dimensions.  However, in obtaining type~I
from type~IIB via the simple orientifold with $\Omega$ in ten
dimensions, the 4--form $A_4$, which is odd under $\Omega$, gets
projected out of the theory and leaves us with no candidate to give
rise to the extra tensors after compactifying.

\item\item{{\bf 2.}} If this was the limit of a smooth type~I 
compactification on $K3$, there should be a limit where we could
enlarge the $K3$ manifold to get an effective ten dimensional theory,
use strong/weak coupling duality to get an $SO(32)$ heterotic theory,
and then shrink $K3$ again, thus deducing six dimensional dual
$SO(32)$ heterotic theories which seem to have extra tensor multiplets
arising apparently at all values of the coupling.

This latter reason by itself is not so unsettling a problem, as we
have learned in recent times to be open to the idea of the appearance
of new structures which cannot be seen perturbatively, but persist at
all values of the coupling. However, it is all too easy to be
satisfied with such an explanation without exploring it further.  In
particular, the deduction that small instantons give rise to new
non--perturbative phenomena\small\ in the heterotic string arose by
considering\edcomm\ a singular limit of a perturbatively well--defined
heterotic object, the solitonic fivebrane instanton\solitons. In some
loose sense then, there was some herald of such peculiar
non--perturbative behaviour in perturbative heterotic string
theory. In this context, the above (bogus) deduction ({\bf 2}) would
lead us to deduce new non--perturbative structures in the $SO(32)$
heterotic string ---the extra tensors--- for the wrong reasons, and we
would not have had a suggestion (analogous to the big fivebrane
instantons) that such phenomena might appear.

So the resolution of the points are as follows:

\item\item{{$\bf 1^\prime$.}} 
The operations of orientifolding and compactifying do not commute, in
general.  The extra tensors arise simply because $A_4$ has not been
projected with $\Omega$, but with $\widetilde\Omega$, which is the
combination of $\Omega$ with a spacetime symmetry under which
(components of) $A_4$ transforms\refs{\atishii,\joe}. It therefore can
survive in the resulting model, contracting with $K3$ two--cycles and
giving rise to self--dual 2--forms in the six dimensional model.  The
choice which was implemented for the operation of $\Omega$ in the
closed string sector implied in ref.\ericme\ was implicitly a
realisation of type\foot{Julie Blum has also computed explicit
realisations of the action of $\Omega$ and $\widetilde\Omega$ in the
closed string sectors of explicit orientifold models.}\ $\widetilde\Omega$.

\item\item{$\bf 2^\prime$.} 
As a result of the choices made, the reduced set of moduli for the
$K3$ surface result in it being impossible\joe\ to blow up the $\IZ_N$
fixed points (for $N{>}2$) to completely smooth manifolds for the
generic models\foot{This is true within the orientifold framework. In
M--theory we will see that the $K3$ we compactify on has the full set
of moduli available.}. This prevents us from being able to glibly
deduce perturbative heterotic models as duals along the lines
indicated above.  The naive deduction via the scenario ({\bf 2}) above
would have led to only an $SO(32)$ (or at best, ${\sl
spin}(32)/\IZ_2$) framework. The phenomenon of extra tensors fits more
naturally into an $E_8{\times}E_8$ setting, which can be either found
in the $E_8{\times}E_8$ heterotic string, or something closely related
to it --- M--theory on an $S^1/\Z_2$ orbifold\petred.

As shown in refs.\refs{\ericjoe,\ericme}\ the open string spectrum
produced massless vector and hypermultiplets resulting in the gauge
content listed in the table below\foot{We list all of the charges of the
hypermultiplets for completeness. The appearances of notation like
`$99$', `$55$' and `$59$' mean massless fields arising from strings
stretched between coincident D--branes of type 9 or 5.}:

\bigskip

\vbox{
$$\vbox{\offinterlineskip
\hrule height 1.1pt
\halign{&\vrule width 1.1pt#
&\strut\quad#\hfil\quad&
\vrule#
&\strut\quad#\hfil\quad&
\vrule#
&\strut\quad#\hfil\quad&
\vrule width 1.1pt#\cr
height3pt
&\omit&
&\omit&
&\omit&
\cr
&\hfil Model&
&\hfil Gauge Group&
&\hfil Charged Hypermultiplets&
\cr
height3pt
&\omit&
&\omit&
&\omit&
\cr
\noalign{\hrule height 1.1pt}
height3pt
&\omit&
&\omit&
&\omit&
\cr
&\hfil $\IZ_2^A$&
& $\eqalign{&99:\quad U(16)\cr &55:\quad U(16)\cr&\phantom{59:\quad}}$&
& $\eqalign{&99:\quad 2\times {\bf 120}\cr&55:\quad 2 \times {\bf 120}\cr
&59:\quad ({\bf 16,16})}$ &
\cr
height3pt
&\omit&
&\omit&
&\omit&
\cr
\noalign{\hrule}
height3pt
&\omit&
&\omit&
&\omit&
\cr
&\hfil $\IZ_3^A$&
& $\eqalign{ 
99:\quad U(8)\times SO(16)\cr
}$&
&  $\eqalign{
99:\quad  {\bf (28,1)}; \,\,{\bf (8,16)}\cr
}$&
\cr 
height3pt 
&\omit& 
&\omit& 
&\omit&
\cr
\noalign{\hrule }
height3pt 
&\omit& 
&\omit& 
&\omit&
\cr
&\hfil $\IZ_4^A$&
& $\eqalign{ 
&99:\quad U(8)\times U(8)\cr
&55:\quad U(8)\times U(8)\cr
&\phantom{59:\quad}}$&
& $\eqalign{
&99:\quad {\bf (28,1)};\,\,\,{\bf (1,28)};\,\,\,{\bf (8,8)}\cr
&55:\quad {\bf (28,1)};\,\,\,{\bf (1,28)};\,\,\,{\bf (8,8)}\cr
&59:\quad ({\bf 8,1;8,1});\,\,\,({\bf 1,8;1,8})}$ &
\cr
height3pt 
&\omit& 
&\omit& 
&\omit&
\cr
\noalign{\hrule } 
height3pt 
&\omit& 
&\omit& 
&\omit&
\cr
&\hfil $\IZ_6^A$&
& $\eqalign{ 
&99:\quad U(4)\times U(4)\times U(8)\cr
&55:\quad U(4)\times U(4)\times U(8)\cr
&\phantom{59:\quad}}$&
& $\eqalign{
99:\quad &{\bf (6,1,1)};\,\,\,{\bf (1,6,1)}\cr
&{\bf (4,1,8)};\,\,\,{\bf (1,4,8)}\cr
55:\quad &{\bf (6,1,1)};\,\,\,{\bf (1,6,1)}\cr
&{\bf (4,1,8)};\,\,\,{\bf (1,4,8)}\cr
59:\quad &{\bf (4,1,1;4,1,1)}\cr
&{\bf (1,4,1;1,4,1)}\cr
&{\bf (1,1,8;1,1,8)}}$ &
\cr 
height3pt 
&\omit& 
&\omit& 
&\omit&
\cr
\noalign{\hrule } 
height3pt
&\omit& 
&\omit& 
&\omit&
\cr
&\hfil $\IZ_4^B$&
&\hfil {\rm ---}&
&\hfil {\rm ---}&
\cr
height3pt 
&\omit& 
&\omit& 
&\omit&
\cr
\noalign{\hrule } 
height3pt 
&\omit& 
&\omit& 
&\omit&
\cr
&\hfil $\IZ_6^B$&
& $55:\quad U(8)\times SO(16)$&
& $55:\quad {\bf (28,1)};\,\,\,{\bf (8,16)}$ &\cr
}
\hrule height 1.1pt}
$$
}

It is interesting to see how the potential chiral anomalies of these
models are cancelled. In particular, the irreducible $\Tr R^4$ and
$\Tr F^4$ terms  vanish, the former by virtue of each model
solving the anomaly equation
\eqn\anomaly{n_H-n_V=244-29n_T,} where $n_H, n_V$ and $n_T$ are the 
number of hyper--, vector-- and extra tensor--multiplets in a model
(perturbative heterotic string theory has $n_T{=}0$). The remaining
anomalies are cancelled by a generalisation of the Green--Schwarz
mechanism\refs{\green,\sagnotti,\joetc}, where all of the tensors in
the model come into play, having one--loop couplings to the gauge and
gravitational sectors, together with their field strengths being
modified from the naive form to one in which gravitational and gauge
Chern--Simons forms appear in the standard way.

Before moving on, let us recall a few things easily noticed about the
models\ericme.  Notice that the $\IZ_3^A$ model is isomorphic to the
$\IZ_6^B$ model. The map between them is the aforementioned
$T_{6789}$--duality, which exchanges D9-- with D5--branes. All of the
other models presented in the table are self--dual under this
operation.  Of these, the $\IZ_4^B$ model has no D--branes and thus no
gauge sector. It is a purely closed string model.

The remaining models, $\IZ_2^A, \IZ_4^A$ and $\IZ_6^A$, are all
similar in some sense (which will be the main focus of this paper):

\item\item{\bf a.} They all contain 32 of both D9-- 
and D5--branes, which get exchanged under $T_{6789}$. The associated
gauge groups and hypermultiplets get exchanged under this operation.

\item\item{\bf b.} They all have gauge groups of the same rank, 
which are successively smaller subgroups of the original
$U(16){\times}U(16)$ as $N$ gets larger.

\item\item{\bf c.} It was learned that away from fixed points, 
the D5--branes were constrained by the consistency conditions to move
as a single unit made of $2N$ D5--branes, forming a `dynamical
fivebrane' which it is tempting to identify with the `small instanton'
unit in a dual heterotic model.  Indeed, they give rise to the same
enhanced gauge groups in bulk as the basic small instanton
example\small, and various patterns of enhancements as they settle on
fixed points\refs{\ericjoe,\ericme}.

\item\item{\bf d.} The numbers  of the  dynamical fivebranes  available in 
each model ($\IZ_2^A, \IZ_4^A$ and $\IZ_6^A$), are 8, 4 and 2
respectively. Adding these to the numbers of extra tensors in each
model, which are 0, 4 and 6 respectively, always results in the number 8.

These features should mean something in the final analysis, and we
can make some guesses as to what they might suggest:

\item\item{$\bf a^\prime$.} In the case of the $\IZ_2^A$ model, a 
relation to a heterotic model has been demonstrated in ref.\joetc.
For a particular arrangement of D5--branes which we will discuss later
(and inclusion of Wilson lines) it is a realisation of the
$E_8{\times}E_8$ heterotic string compactified on $K3$ with instanton
embedding (12,12). The perturbative $T_{6789}$--duality, which acts
non--trivially on the vector-- and hyper--multiplets coming from the
D9-- and D5--brane sectors, gets mapped to the heterotic/heterotic
duality map, which acts non--trivially on the perturbative vector--
and hyper--multiplets to exchange them with those appearing
non--perturbatively due to small ${\sl spin}(32)/\IZ_2$ instantons.

In the blow--down limit of the $K3$ surface in the orientifold, the
instanton number 24 is distributed amongst the $8$ small instantons
and the 16 fixed points of the orbifold which are the blowdown of
Eguchi--Hanson spaces\eguchi\ ${\cal E}_2$.

So in the other models, when we find dual realisations of them, we can
hope to find an understanding of what the operation $T_{6789}$ maps
to.

\item\item{$\bf b^\prime$.} 
In all of the models we will choose the special case where the gauge
group is generally completely broken. It will be these cases which
have a relation to the M--theory models which we construct.

\item\item{${\bf c^\prime}$ $\&$ ${\bf d^\prime}$.} 
There is some description of a `parent model' in which a single type
of object, of which there are 8 in total, are present for reasons of
charge cancellation, or some other (perhaps topological) reason.  We
can move between different parts of the moduli space of this parent
model, and in a dual orientifold setting, we realize one of models
$\IZ_2^A,\IZ_4^A$ or $\IZ_6^A$, depending upon the details.

Let us try to  discover the nature of  this parent model.

\newsec{M--Theory}
M--theory is the first setting in which we shall try to fit the rest
of our models. As discussed before, it is not necessary to appeal to
M--theory to understand the $\IZ_2^A$ model, but when we try to
understand the extra tensors in the other models it is very natural.

The strong coupling limit of $E_8{\times}E_8$ heterotic string theory
in ten dimensions has been shown\petred\ to be a simple `orbifold' of
the eleven dimensional M--theory, where the eleventh dimension is
placed on an orbifold $S^1/\IZ_2$. Not much is known about orbifolds
of this still unknown theory, but whatever happens should of course
not contradict results we know to be true in string and field
theory. In this spirit, the authors of ref.\petred\ showed that the
ten dimensional spacetime living at each end of the line segment
resulting from the orbifold should give rise to an $E_8$ gauge group,
which give rise to the $E_8{\times}E_8$ of the heterotic string in the
weak coupling limit (when the size of the $S^1/\Z_2$ goes to zero).

A consistent heterotic string compactification on $K3$
requires\schwarz\ a choice of a gauge bundle of instanton number
24. This choice can be split between the factors of the gauge group in
a way labeled by the integer $n$, placing instanton number $12{-}n$ in
the first $E_8$ and $12{+}n$ in the other.  These $(12{-}n,12{+}n)$
embedding models have been discussed extensively in the literature
recently. The spectrum resulting from this embedding is determined by
an index theorem\schwarz\ which yields the number of hypermultiplets
in the {\bf 56} of the resulting $E_7$ gauge group (for a choice of
$SU(2)$ instanton bundles). In particular, for $n{=}0,2$ we have
enough {\bf 56}'s, to allow us to break the gauge group completely, by
sequential use of the Higgs mechanism. (As mentioned before, the cases
$n{=}0$ and 2 turn out to be related to one another.) This results in
a number of uncharged hypermultiplets, which must equal 244, by the
anomaly equation \anomaly.

The special case of the $\IZ_2^A$ model\ericjoe\ related to this model
is as follows: It is possible to arrange the D5--branes (and by
$T_{6789}$-duality, introduce Wilson lines) in such a way as to break
the gauge symmetry carried by the D5--branes (D9--branes)
completely. This is done by placing two D5--branes at each of the 16
$\IZ_2$ fixed points. Naively, this yields a gauge group $U(1)^{16}$,
according to the analysis of ref.\ericjoe, but the work of ref.\joetc\
shows that such $U(1)$ groups are broken. This again results in no
vectors and a number of uncharged hypermultiplets, which must equal
244, by the anomaly equation \anomaly.

This unique configuration has at least one other interesting property
of note. The $\IZ_2$ fixed points carry $-2$ units of charge in
D5--brane units. This is one way to determine that there are 32
D5--branes ($={}8$ dynamical fivebranes) available in the problem, for
consistency. By the arrangement above, not only is charge cancellation
satisfied in the compact space, but further to that it is satisfied
locally. This statement applies (by supersymmetry) to the cancellation
for the NS-NS sector, implying in particular that the dilaton has been
cancelled locally. This in turn assures us that a perturbative
heterotic dual can be found\edjoe. Other configurations where the
charges were not cancelled locally would have regions where the dilaton
could approach values corresponding to regions of strong coupling in
any dual picture, thus spoiling a strong/weak coupling duality
construction.
 
Notice also that as the minimal fivebrane object allowed to move in
the bulk ({\it i.e.,} off the fixed points) is a collection of 4
D5--branes, this configuration has no flat directions corresponding to
un--Higgs--ing back to the generic gauge groups found in ref.\ericjoe,
as there are half--fivebranes on each distinct point. At this level,
there is nothing to rule out the possibility of another Higgs--ing
route to another branch with different gauge groups. One such branch
available is the $E_7$ one of the $K3$ compactified $(12,12)$
heterotic model to which this has been shown to be equivalent\joetc.

Something new arises when we realize that there are other ways of
constructing consistent M--theory
compactifications\refs{\sunil,\edfive}. The consistency condition can
be thought of as a charge cancellation condition for the 6--form
potential in M-theory, obtained by dualising the 3--form. (This is
related of course to the R-R 6--form charge cancellation condition of
the orientifold model). More properly, there is the usual ten
dimensional correction to the field strengths from both the geometry
and gauge sectors in order to cancel the anomaly. The eleven
dimensional theory is anomaly free, except at the ten dimensional
boundaries of the orbifold interval. The boundaries therefore act as
sources of the anomaly equation in eleven dimensions. The objects
which live on the ten dimensional subspaces which we eventually refer
to as instantons in the string theory carry unit charge.  In addition
to the conventional placing of instantons, we can also distribute the
charge amongst some M--theory fivebranes, which naturally carry (unit)
charge also\edfive.  So when we compactify on $K3$, which supplies the
opposite charge equal to 24 via geometry, we have an equation:
\eqn\cancel{n_1+n_2+n_T=24,}
where $n_1$ and $n_2$ are the instanton contribution from the ten
dimensional spacetimes at the end of the $S^1/\IZ_2$ interval, while
$n_T$ is the number of M--theory fivebranes present. Of course, now we
can see the dual nature of the term `fivebranes' from our M--theory
point of view.  When they are living at the ten dimensional fixed
points, they are the more conventional fivebranes which we recognized
as the fully dressed string theory instantons (made of collections of
D5--branes), and away from the the fixed points they are the eleven
dimensional M--theory theory fivebrane.  Where ever they happen to be
in the higher dimensions, for the purposes of this consistent $K3$
compactification, the fivebranes must be transverse to the compact
space. This means that their world volume is aligned with the
non--compact directions $\{X^0,\ldots,X^5\}$. 

It is no accident that
we denoted the number of fivebranes in the interior of the interval in
$X^{10}$ by $n_T$. This is because the contribution of the fivebrane's
worldvolume to the six dimensional spectrum is one tensor multiplet of
${\cal N}{=}2$ supersymmetry\foot{Here we denote the spacetime
transformation properties of states by their dimension as a
representation of the six dimensional little group
$SU(2){\times}SU(2)$.}\edfive: ${\bf (3,1)}{+}5{\bf (1,1)}$. In our
${\cal N}{=}1$ context, this is a contribution of a tensor multiplet
$\bf(3,1){ +}(1,1)$ and a hypermultiplet $4{\bf (1,1)}$, precisely
what the closed string sector supplies from the $\IZ_N$ orientifold
fixed points ($N{\neq2}$), as computed in ref.\ericme.

We are now in good shape to begin trying to understand how our other
orientifold models might fit into the M--theory picture.  Let us first
try to understand precisely what models we wish to consider. In the
case of the models $\IZ_N^A$ which we are considering (for $N$ even),
there is a more complicated orientifold fixed point structure, as
discussed in ref.\ericme. However, there is one simplifying
observation\ericme. Regardless of the structure of the fixed points,
in each case there is only one type of fixed point which is
responsible for acting as a source of ({\sl untwisted}) R-R 6--form
charge (under discussion here) and that is the $\IZ_2$ fixed point,
the spacetime manifestation of the element $\Omega R$ in the
orientifold group. ($R$ denotes spacetime reflection in the
$X^6,X^7,X^8,X^9$ directions, which forms a $\IZ_2$ subgroup of all
the models under consideration here).  There are always 16 of these
$\IZ_2$ fixed points , which is why there are always 32 D5--branes. In
other words, each such fixed point carries $-2$ units of D5--brane
charge, as before\foot{The difference between the models arises when
we realize that the $\IZ_N$ orbifold acts by grouping $N$ pairs of
D5--branes into dynamical fivebrane units, making fewer of them
available as $N$ increases. They are paired because\ericjoe\ of the
presence of $\Omega$.}.

We can therefore construct in each of these orientifold models the
same special configuration yielding local cancellation of the untwisted
R-R 6--form charge. Arguments analogous to those in ref.\joetc\ ensure
that the naive $U(1)^{16}$ gauge group is again broken completely,
leaving only uncharged hypermultiplets and the tensor multiplets. The
spectrum is then easy to determine for each model, and is listed below
in summary:

\bigskip
\vbox{
\eqn\tablethree{
\vbox{\offinterlineskip
\hrule height 1.1pt
\halign{&\vrule width 1.1pt#
&\strut\quad#\hfil\quad&
\vrule#
&\strut\quad#\hfil\quad&
\vrule#
&\strut\quad#\hfil\quad&
\vrule width 1.1pt#\cr
height3pt
&\omit&
&\omit&
&\omit&
\cr
&\hfil Model&
&\hfil \vbox{\hbox{$n_H$}}&
&\hfil \vbox{\hbox{$n_T$}}&
\cr
height3pt
&\omit&
&\omit&
&\omit&
\cr
\noalign{\hrule height 1.1pt}
height3pt
&\omit&
&\omit&
&\omit&
\cr
&\hfil $\IZ_2^A$&
&\hfil 244&
&\hfil 0&
\cr
height3pt
&\omit&
&\omit&
&\omit&
\cr
\noalign{\hrule}
height3pt
&\omit&
&\omit&
&\omit&
\cr
&\hfil $\IZ_4^A$&
&\hfil 128&
&\hfil 4&
\cr
height3pt 
&\omit& 
&\omit& 
&\omit&
\cr
\noalign{\hrule } 
height3pt 
&\omit& 
&\omit& 
&\omit&
\cr
&\hfil $\IZ_6^A$&
&\hfil 70&
&\hfil 6&
\cr 
height3pt 
&\omit& 
&\omit& 
&\omit&
\cr
}
\hrule height 1.1pt}
}
}

Turning back to the M--theory interpretation, if we interpret the
tensors as coming from pushing fivebranes out into the interior of the
eleven dimensional interval, then we are left with the task of
distributing the remaining instanton number amongst the $E_8$ gauge
groups when we compactify. We have certain constraints on how we can
distribute if we wish to find our models. We must find configurations
which give us a spectrum which can be Higgs--ed away to nothing, which
is the gauge content of all of our models, as chosen above. Embedding
instanton number~8 into $E_8$ for an $SU(2)$ gauge bundle leaves gauge
group $E_7$ with 2 hypermultiplets in the $\bf 56$, by the index
formula.  This is not enough matter to break the group completely,
leaving an unbroken $SO(8)$. Smaller amounts of instanton number
produce larger unbroken gauge groups, and so we must consider
instanton number greater than 8. Instanton numbers 12, 11 and 10
produce 4, $3\half$ and 3 $\bf 56$'s respectively, and are known to
have sequential Higgs--ing routes which lead to completely broken
gauge groups. Instanton number 9 is interesting, however. By the index
formula, it produces $2\half$ $\bf 56$'s, which is (naively) enough to
break the $E_7$ gauge group completely (since $140>133$). However, it
seems that the Higgs--ing route to a completely broken gauge group has
not yet been found. However, as pointed out in ref.\hethet, failure to
find a Higgs--ing route does not rule out the existence of such a
branch of moduli space. We shall assume that it exists, and our
motivation will simply be that it fits all the available data of our
models.

With this information in mind, we have the following candidate
arrangements for our models. $\IZ_4^A$ has $(n_1,n_2)=(10,10)$ or
$(11,9)$, while $\IZ_6^A$ has $(9,9)$ as the only possibility.  Given
the data and techniques we have to work with, we have no way to decide
between the two choices\foot{The choice (10,10) seems more
aesthetically pleasing, as it matches the (9,9) and (12,12) of the
other models.}\ for $\IZ_4^A$. However, there is the possibility that
the choices are related in the same way that the (12,12) and (10,14)
models are related.

Another way to look at things is to examine how the spectra we found
above match what we would naively expect from the instanton moduli
space.  For large instanton number $n$, the index formula gives us an
expression for the dimension of the moduli space ${\cal M}_n$ of $E_8$
instantons on $K3$:
\eqn\moduli{{\rm dim} {\cal M}_n(E_8)=120n-992.}

Putting in the numbers for each case $(n_1,n_2)$, and dividing the
result by 4 to get the number of hypermultiplets, we find the result
224 for $\IZ_2^A$, 104 for $\IZ_4^A$ and 44 for $\IZ_6^A$. This is
consistent with the above table of hypermultiplets if we add 20
hypermultiplets in each case for $K3$ gravitational moduli, and $n_T$
hypermultiplets in each case corresponding to the positions\foot{The
movement of the M--theory fivebranes along the interval does not
correspond to a true modulus of the theory. This is in line with the
fact that the associated scalar is in a tensor multiplet and not a
hypermultiplet.}\ of the fivebranes in the $K3$.

We also wished to gain some insight into the importance of the
number~8 in this setting. Now we see that it is simply the total
number of M--theory fivebranes in the problem. In each of the
orientifold models, the $K3$ manifold as an orbifold acts as a source
of $-24$ units of 6--form charge in equation \cancel, (from geometry)
and $+16$ units from instanton charge, localised in the 16 $\IZ_2$
fixed points which are present. The other fixed points do not
contribute to the counting as they have no 6--form charge. We expect
that this distribution of instanton number is preserved in going to
the M--theory compactification on smooth $K3$. So there, in order to
cancel the remaining $-8$ units of charge, we have to introduce 8
fivebranes.

The order $N$ of the spacetime symmetry of the orientifold models
determines in M--theory how many fivebranes live on the
ten--dimensional spacetime of an $S^1/\Z_2$ orbifold fixed point,
playing the role of string theory instantons/mulitple D5--branes,
leaving the rest on the $X^{10}$ orbifold interval.

We have thus found a natural setting in which to place the orientifold
models $\IZ_2^A,\IZ_4^A$ and $\IZ_6^A$, which `explains' their
similarities and differences\foot{We should also mention that an
M--theory realisation of a model closely related to the $\Z_4^B$
model\atishi\ was worked out in ref.\senM. The eight extra tensors are
produced by eight M--theory fivebranes, required by the presence of
fixed points of an orbifold of smooth $K3$ by the Enriques
involution.}. We expect that these models are all connected by a phase
transition (from the six dimensional point of view) occurring when a
fivebrane detaches from the ten dimensional world volume and goes into
the bulk of the eleven dimensional
spacetime\refs{\ganor,\edphase\duffphase}. This process lies naturally
outside the description of perturbative heterotic string theory, which
is why we had such difficulty interpreting the duals of the models
inside a string theory framework (other than the type~IIB orientifold
framework).

We should note here that if we really want to cling to the idea of
string theory, we can naively take the small $S^1/\Z_2$
limit\foot{Well, something more like a multiple scaling limit where we
take the interval size to zero while holding the positions of the
$n_T$ fivebranes away from the interval's edges.}\ of the M--theory
construction to recover the $E_8{\times}E_8$ $K3$ compactified
heterotic string and interpret the $n_T$ extra tensors as `new
non--perturbative data', as we suggested at the outset\foot{There must
be a $T$--duality relationship (generalising the one in ref.\joetc)
between the resulting $E_8{\times}E_8$ string theory and the $SO(32)$
one, realising the scenario ($\bf 2$) described in the first part of
section~2. Although probably complicated, due to the presence of the
extra tensors, it should exist, given the fact that the $A$--type
orientifold models are locally $SO(32)$ type~I theory. We thank Joe
Polchinski for pointing out this possibility to us.}. The $T_{6789}$
self duality of the orientifold models would then induce the action of
a generalisation of the heterotic/heterotic duality for these string
theories. The explicit construction of the fundamental strings
which would get exchanged under this duality as solitons in our
orientifold theories should be straightforward. One such string is
simply the D1--brane, which we can arrange to live in the non--compact
directions. The other is a D5--brane wrapped around the $K3$, also
giving a strings in the non--compact directions, $T_{6789}$--dual to
the D1--brane\joetc.

\newsec{Some Four Dimensional  Dualities}
Let us compactify our M--theory models further on a torus $T^2$. Now
we are in a four dimensional setting, with ${\cal N}{=}2$
supersymmetry. Recall that the six dimensional tensor multiplets {\sl
as well as} the vector multiplets give rise to four dimensional vector
multiplets. Meanwhile hypermultiplets map to hypermultiplets, and the
compactification on the extra torus gives us an extra four vector
multiplets for gauge group $U(1)^4$.

So now we have three models with spectra which are given by the same
number $n_H$ of hypermultiplets as given in \tablethree, with
$n_V{=}n_T{+}4$.

This situation resembles something else that we have seen before. In
ref.\shamit, a number of four dimensional ${\cal N}{=}2$ vacua were
constructed by compactifying the heterotic string on $K3{\times}T^2$,
with a special choice of gauge bundle. In addition to embedding
instanton numbers $(n_1,n_2)$ into each $E_8$, there was an embedding
of $n_T$ units into the non--Abelian gauge group obtained by placing
the torus $T^2$ at a special point in its moduli space. Consistency
was achieved by ensuring that the sum \cancel\ was satisfied.

In the resulting four dimensional setting, there is no way of knowing
whether the spectrum has originated from such a $K3{\times}T^2$
heterotic string compactification such as that carried out in
ref.\shamit, or a $K3{\times}T^2{\times}S^1/\IZ_2$ M--theory
compactification with fivebranes, as described here. This is
suggestive of a new and interesting duality relationship between the
two, which deserves further exploration. It probably involves a map
between the geometry of the special torus (with its gauge bundle of
instanton number $n_T$) and the eleventh dimension's orbifold interval
(and its $n_T$ fivebranes)

Instead of following that avenue of investigation, there is yet
another duality which is relevant in this setting, which was the
subject of ref.\shamit. The heterotic vacua we have just constructed
in four dimensions are described at strong coupling by a
compactification of type~IIA string theory on Calabi--Yau 3--folds
with the Hodge numbers $h_{2,1}{=}n_H{-}1$ and $h_{1,1}{=}n_T{+}3$,
defining for us three 3--folds $Y_1,Y_2$ and $Y_3$:

\bigskip
\vbox{
\eqn\tablefour{
\vbox{\offinterlineskip
\hrule height 1.1pt
\halign{&\vrule width 1.1pt#
&\strut\quad#\hfil\quad&
\vrule#
&\strut\quad#\hfil\quad&
\vrule#
&\strut\quad#\hfil\quad&
\vrule width 1.1pt#\cr
height3pt
&\omit&
&\omit&
&\omit&
\cr
&\hfil C--Y 3--fold&
&\hfil \vbox{\hbox{$h_{2,1}$}}&
&\hfil \vbox{\hbox{$h_{1,1}$}}&
\cr
height3pt
&\omit&
&\omit&
&\omit&
\cr
\noalign{\hrule height 1.1pt}
height3pt
&\omit&
&\omit&
&\omit&
\cr
&\hfil $Y_1$&
&\hfil 243&
&\hfil 3&
\cr
height3pt
&\omit&
&\omit&
&\omit&
\cr
\noalign{\hrule}
height3pt
&\omit&
&\omit&
&\omit&
\cr
&\hfil $Y_2$&
&\hfil 127&
&\hfil 7&
\cr
height3pt 
&\omit& 
&\omit& 
&\omit&
\cr
\noalign{\hrule } 
height3pt 
&\omit& 
&\omit& 
&\omit&
\cr
&\hfil $Y_3$&
&\hfil 69&
&\hfil 9&
\cr 
height3pt 
&\omit& 
&\omit& 
&\omit&
\cr
}
\hrule height 1.1pt}
}
}

Of course, $Y_1$ is already extremely well known, in this and other
related contexts and is the 3--fold denoted $X_{24}(1,1,2,8,12)$
earlier. The other two are not known to us at the time of
writing. However, we expect that they exist, and furthermore that they
exist as $K3$ fibrations over $\IP_1$, where the size of the base
determines the strength of the heterotic string coupling, a basic
result of heterotic/type~IIA duality\klemm. We will ask some more of
them in the next section.

\newsec{F--Theory}

In the previous section, by compactifying on $T^2$ and invoking four
dimensional heterotic/type~IIA duality, we arrived at type~IIA string
theory vacua. This is amusing, since we started out by orientifolding
type~IIB string theory, and we might wonder whether we can complete
the circle of dualities, perhaps learning more along the way.

Well, the standard route from here to get to type~IIB string theory
would be perhaps to use mirror symmetry, and study a compactification
on the mirrors of the manifolds in
\tablefour. This is not an attractive route for the interests of this
paper for at least two reasons. The first is that we know so little
about the manifolds $Y_2$ and $Y_3$, and so the investigation would be
rather short. The second is that it is will not obviously lead us back
to a six dimensional type~IIB setting to complete the circuit.

The route we wish to take was already deduced in
refs.\refs{\vafaF,\vafamorrison}. Starting with type~IIA string theory
compactified on a Calabi--Yau manifold $X$, one can imagine a limit
(loosely, $X$ has to be large) in which the strong coupling limit
might be captured by M--theory on $S^1{\times}X$. From here, we can
use a number of conjectured dualities to go into almost any direction
we want. For example, we can exploit the fact that $X$ is a $K3$
fibration and try deduce a duality fibre by fibre by wrapping the
M--theory fivebrane on $K3$ eventually deducing again the relation to
heterotic strings.  Another route is to require that $X$ has an
ellitpic fibration, compactify on the resulting torus to nine
dimensions where we can $T$--dualise to type~IIB theory, and from
there go on to F--theory on $X{\times}S^1{\times}S^1$. When we let
this new two torus grow large, we have F--theory compactified on the
3--fold $X$.

F--theory compactified on a
Calabi--Yau 3--fold gives six dimensional ${\cal N}{=}1$ vacua that
are related to heterotic string vacua as first set out in
refs.\refs{\vafamorrison}, by relating it to the $Y_1$ example of 
heterotic/type~IIA and heterotic/heterotic duality.

Following that route we see that there is a natural interpretation in
F--theory of many other four dimensional ${\cal N}{=}2$ vacua existing
as dual heterotic/type~IIA pairs. The associated Calabi--Yau 3--fold
would have to be elliptic, in order to have a six dimensional
interpretation when used as an F--theory compactification. The six
dimensional vacua will contain extra tensors, the number of which is
given by $h_{1,1}(B){-}1$ where $B$ is the base of the elliptic
fibration of the 3--fold\vafamorrison.

So the extra requirement we ask of our manifolds in \tablefour\ is
that they are all elliptic, yielding for us new F--theory backgrounds
with spectra given in \tablethree.

\newsec{A Return to Orientifolds}

Until we learn more about F--theory, we are free to regard it as a new
way of learning about backgrounds for the type~IIB string. In this
sense, it is very much akin to the orientifold technology, and it
would be nice to make something of this.

The torus of the twelve dimensions of F--theory is a geometrisation of
the coupling `constant' of type~IIB theory. On compactifying on an
elliptic manifold to some dimension lower than ten, we are
constructing a compactification of type IIB strings on the base $B$ of
the fibration. For example, in the case of compactification to six
dimension by means of an elliptic Calabi--Yau 3--fold, the base $B$ is
not itself Calabi--Yau, and so we have naively obtained a sick IIB
background. However, there are sevenbranes present in the problem,
their positions given by the locations in the base $B$ where the torus
fibre degenerates. The presence of the sevenbranes completes the
consistency requirement for yielding a IIB background.

That the sevenbranes are present is very natural. The torus fibre
parameterizes the IIB coupling via the modulus
$\tau(z_i){=}A_0{+}e^{-{\Phi\over 2}}$ where $\Phi$ is the dilaton and
$A_0$ is the R-R 0--form (scalar) potential. The $z_i$ are coordinates
on the base $B$ over which the torus $T^2$ is fibred. The $SL(2,\IZ)$
self duality symmetry of the type~IIB string acts on the torus. Recall
that the sevenbrane is the natural object in the theory carrying
electric charge of (the field strength of) $A_8$ and hence magnetic
$A_0$ charge. The degeneration of the torus at positions on the base
is simply a signal the presence of a magnetic source of $A_0$, the
sevenbrane\vafaF.

Consistency requires a certain number of sevenbranes to be present in
order to correctly cancel the $A_0$ charge. We have heard this story
three times before in this paper. Once in the context of orientifolds
and D--branes, once in the context of heterotic string theory and
consistent $K3$ compactifications with instantons and once more in
M--theory in the context of combining $K3$ compactifications with
fivebranes. We discussed the relations between them. We should
therefore expect that this charge cancellation is once again the
orientifold charge cancellation in disguise and we shall see that it
is.

In ref.\senF, a precise relation between the construction of F--theory
on a smooth elliptic $K3$ manifold and a type~IIB orientifold was
made. This construction enabled a more precise demonstration that the
resulting background was dual to heterotic string compactified on
$T^2$.  The $K3$ is fibred as $T^2$ over a base $\IP_1$. Generically,
the fibre degenerates at 24 positions, implying that there are that
many sevenbranes in the problem, their worldvolumes aligned with the
uncompactified eight dimensions with a point--like intersection~on the
$\IP_1$. At this stage, it is not quite true to say that this is a
type~IIB string theory background obtained by compactifying on $\IP_1$
and including 24 branes, as stressed in ref.\vafaF. Clearly, the
string theory description could not be perturbative, as the coupling
is varying all over the $\IP_1$.

It was demonstrated in ref.\senF\ however, that a limit could be
approached where the torus parameter $\tau(z)$ does not vary smoothly
over the base but is constant with the possibility of phases at a
finite number of positions.  In that limit, the type~IIB string theory
description can be made to work. It is a type of $\IZ_2$ orientifold
of the theory on a torus $T^2$. The orientifold requires 32 D7--branes
to be in the problem, for charge cancellation. The four fixed points
of the orientifold have charge $-8$ in D7--brane units and the charge
cancellation can be carried out locally by placing 8 D--branes at each
of the points.

As mentioned before, F--theory on the 3--fold $Y_1=X_{24}(1,1,2,8,12)$
is another realisation of the (12,12) $K3$ heterotic compactification,
which in turn has a realisation as the $\IZ_2^A$ orientifold model.

It is natural to wonder if we can directly find a relation between the
F--theory compactification and the orientifold, along the lines of
ref.\senF.  It is easy to see the relation if we begin by T--dualising
along two of the directions of the orientifold four torus. Let us
choose directions $X^6$ and $X^7$.  Recalling that T--duality
exchanges Dirichlet and Neumann boundary conditions, we see that the
D5--branes get converted to D7--branes with world volumes located
along direction $X^\mu$ for $\mu{\in}\{0,1,2,3,4,5,6,7\}$, while the
D9--branes get converted to D7--branes located along the directions
given by $\mu{\in}\{0,1,2,3,4,5,8,9\}$ directions.

The presence of the D7--branes is already heartening, in view of their
natural occurrence in F--theory.  Examining the details of the duality
more carefully, we see that we are carrying out an orientifold of the
torus $T^2{\times}T^2$ with group
\eqn\orientifold{G=\{ 1, R_{6789}, \Omega
R_{67}(-1)^{F_L},\Omega R_{89}(-1)^{F_L}\}.} 
Geometrically the $\IZ_2$
actions on the tori are given by the reflections
$R_{67}$ and $R_{89}$ respectively. Of course,
$R_{6789}{=}R_{67}{\cdot}R_{89}$. The appearance of $(-1)^{F_L}$,
where $F_L$ is the left fermion number can be traced to the action of
$T$--duality as an action of world sheet parity which is restricted to
only the right or left movers\refs{\dnotes,\atishii}. Our conventions
are such that the action is on the left.

We should expect that the orientifold torus $T^2{\times}T^2$ is
related to the base $\IP_1{\times}\IP_1$ of the elliptic manifold of
the smooth description in the same way that the orientifold torus
$T^2$ was related to the base $\IP_1$ in the eight dimensional
example\senF.  Let us see how this works.

In the eight dimensional example, a Weierstrass representation was
used for the elliptic fibration of $K3$ as a torus $T^2$ over the
sphere $\IP_1$.  There is a standard extension to an elliptic
fibration over $\IP_1{\times}\IP_1$:
\eqn\weir{y^2=x^3-f(z_1,z_2)x-g(z_1,z_2),} giving the 3--fold 
$X_{24}(1,1,2,8,12)$ when $f$ and $g$ are polynomials of degree $8$
and $12$ in the $z_i$, which are coordinates of the $\IP_1$'s.
Counting parameters, one can verify existence of the 243 complex
deformations, after taking into account the rescaling freedom (on the
affine coordinates $x$ and $y$) and the $SL(2,\IC){\times}SL(2,\IC)$
symmetry.  The symmetry exchanging the two $\IP_1$'s is
also manifest, translating into heterotic/heterotic duality as
mentioned previously\vafamorrison.

At any point on the base $B$, the $K3$ fibration is also clearly
visible, the fibre itself being elliptic. The worldvolume of the
sevenbranes on the base $B$ is given by the vanishing of the
discriminant
\eqn\disc{\Delta=4f^3-27g^2,} 
an equation which has 24 solutions, generically.

In the work of ref.\senF, a special point in the moduli space of the
$K3$ was chosen such that the modular parameter of the torus fibre
$\tau(z)$ (determined from the above representation implicitly in
terms of the elliptic $j$--function) is independent of $z$. Here, $z$
is the coordinate of the $\IP_1$ base of the $K3$ fibration. We can
take it to be either $z_1$ or $z_2$ here, and the analysis will go
through with the other $\IP_1$ remaining a spectator. In this limit,
the discriminant $\Delta$ takes a form which indicates that the 24
sevenbranes have coalesced into four groups of six coincident branes,
located around four fixed points $z_1,z_2,z_3,z_4$. The parameter
$\tau$ is constant over the base, with a non--trivial $SL(2,\IZ)$
monodromy around each of the points.

Furthermore, the metric of the base was computed in this limit, and it
turns out to be globally a flat geometry, together with a deficit
angle of $\pi$ at each of the four special points where the
sevenbranes are located.

The orientifold interpretation of this scenario\senF\ is that the
monodromy around the points is $(-1)^{F_L}\Omega$ while the geometry
of each point is that of a $\IZ_2$ fixed point of a spacetime
refection action on a two torus.  In other words, the orientifold
group element is $\Omega R_{89}(-1)^{F_L}$, which we see appearing in
the $T_{67}$ dual of the $\IZ_2^A$ model defined by the orientifold
group in eqn. \orientifold.  In completing the rest of such an
orientifold model, tadpole cancellation would require the addition of
open string sectors in the form of 32 D7--branes. A computation would
reveal that there are $-8$ units of D7--brane charge on each of the
four fixed points and the condition for local cancellation of the
charge (associated to the field strength of the R-R 8--form $A_8$,
dual to $A_0$, part of the coupling $\tau$) is to group them into four
groups of 8 coincident branes\foot{The attentive reader may have by
now noticed a discrepancy of a factor of two between the counting of
ref.\senF\ and the counting here. All is well, for ref.\senF\ counts a
D--brane and its mirror as one object, while we count them as
two.}. This final configuration matches that of the special F--theory
configuration.

So far, we have simply forgotten about the other component of the
base, the $\IP_1$. It is simply brought into the discussion by
carrying out the same procedure again, this time forgetting about the
first $\IP_1$. As the base is simply a product of the $\IP_1$'s there
is nothing lost in this piecewise approach.  This time we end up with
the same orientifold story, with coordinates $\mu{\in}\{6,7\}$ instead
of $\{8,9\}$. Having thus deduced the presence of the operation
$\Omega R_{67}(-1)^{F_L}$, then we are forced by closure to have
$R{=}R_{6789}$ in the orientifold group too.

Thus we see that we have recovered the orientifold group \orientifold\
we deduced from $T_{67}$--dualising the $\IZ_2^A$ model, forging
another connection between F--theory on $X_{24}(1,1,2,8,12)$ and
heterotic/heterotic duality.

The next task (beyond the scope of this paper) would be to carry out
the same procedure for the models $\IZ_4^A$ and $\IZ_6^A$ starting
with the manifolds $Y_2$ and $Y_3$. The procedure of $T_{67}$
dualising is obvious. However, the orientifold action on the
$\mu{\in}\{6,7,8,9\}$ torus $T^4$ will not be factorisable, as one
might expect from the fact that the base manifold of the associated
(conjectured) elliptic fibrations of the 3--folds $Y_2$ and $Y_3$ in
\tablefour\ is unlikely to be a trivial product. This is especially
true since it must have $h_{1,1}$ large enough to yield non--zero
$n_T$ in the final six dimensional spectrum.

Despite that complication, we would expect to see an involution of the
base $B$ of the manifolds $Y_2$ and $Y_3$, which provide a geometrical
realisation of the exchange symmetry $T_{6789}$. This would be the map
which generalizes the heterotic/heterotic duality map of the simpler
model. It is worth exploring what the consequences of such a map would
be for the M--theory compactifications and related theories.

\newsec{Conclusions}
We have come full circle in our  exploration of a chain of dualities. We
started with the orientifold models of refs.\refs{\ericjoe,\ericme},
and related special cases of them to M--theory compactifications to
six dimensions on $K3{\times}S^1/\IZ^2$ with extra fivebranes. This 
extends the correspondence found in ref.\joetc\ to the $K3$
compactified $(12,12)$ $E_8{\times}E_8$ heterotic string for the case
with no extra tensors.

Going  to four dimensions via the torus $T^2$, we noted that
the spectra found there can be viewed as having a dual origin, either
from M--theory with extra fivebranes or from heterotic strings with
instantons embedded in the gauge group arising from placing the torus
at a special point. That relationship deserves further exploration.

We used four dimensional heterotic/type~IIA
duality to deduce what properties two new (to the authors) Calabi--Yau
3--folds would need to have to be associated with the original
orientifold models.

Being in type~IIA string theory it was natural to try to seek a
type~IIB relationship, which led us to F--theory. From there, for the
model for which we have all of the data on the Calabi--Yau 3--fold, we
were able to simply extend the ideas of ref.\senF\ down a further two
dimensions to recover a $T$--dual of the orientifold model we first
started with. We have thus found a direct relationship to F--theory's
smooth description. We expect that this works for the other two
orientifold examples we considered in this paper.

This completes our instructive tour of the duality circuit,
illustrating many links between ideas. Many of the links were
organised by (or have a simple interpretation in) the framework of M--
and F--theory, the parent theories from which all string theories seem
to originate.

\bigskip\bigskip{\bf Note Added:}

While preparing this manuscript for publication, ref.\philip\
appeared, in which related work is presented.

\bigskip
\medskip

\noindent
{\bf Acknowledgments:}

\noindent
We would like to thank Julie Blum for useful conversations. We would
also like to thank Joe Polchinski for useful discussions and comments
on the manuscript. This work was supported in part by the National
Science Foundation under Grants PHY91--16964 and PHY94--07194.
\listrefs
\bye